\documentclass[preprint,amsmath,amssymb]{revtex4}

\usepackage{graphicx}
\usepackage{dcolumn}
\usepackage{bm}

\begin{document}


\title{A pumped atom laser}

\author{N. P. Robins}
 \email{nick.robins@anu.edu.au}
 \author{C. Figl}
\author{M. Jepessen}
\author{G. R. Dennis}
\author{J. D. Close}%

\affiliation{Australian Centre for Quantum Atom Optics, Physics Department, Australian National University}%

\date{\today}

\begin{abstract}
The invention of the optical laser, a bright, 
pumped source of coherent photons, has revolutionized optics and precision measurement, led to global high bandwidth communication and enabled non-linear optics, quantum optics, and experimental investigations of quantum information, quantum computing and quantum cryptography.
The atom laser, a bright, coherent matter wave, holds similar promise for precision measurement and will open new avenues for fundamental tests of quantum mechanics \cite{craig}.  Atom laser beams are derived from the exotic state of matter known as a Bose-Einstein condensate (BEC), first produced in dilute alkali gases in 1995 \cite{cornell,ketterle2}.  These incredibly delicate states exist at almost zero temperature and exhibit many of the properties of optical lasers. Despite significant experimental efforts, no method has been demonstrated to allow continuous and irreversible replenishment of a trapped Bose-Einstein condensate while simultaneously producing a free, coherent atom beam.    
It has taken a decade since the production of the first pulsed atom laser\cite{ketterle1} to surmount a number of serious theoretical and technical hurdles.  In this Letter, we show how we have solved these challenges, demonstrating that while continuously output-coupling an atom laser beam we can simultaneously and irreversibly pump new atoms from a physically separate cloud into the trapped Bose-Einstein condensate that forms the lasing mode. 
 \end{abstract}
\maketitle 
Originally demonstrated by the MIT group \cite{ketterle1}, atom lasers were produced by coherently releasing wave-packets from a BEC.  Pulsed radio frequency (RF) waves changed the magnetically trapped state of the atoms to an untrapped state, and the atoms fell out of the trap under gravity.
Later, Esslinger et al. used this RF method to produce long pulse atom laser beams\cite{esslinger}.  
Recently, multi-photon Raman transitions have been used to produce atom lasers with superior transverse modes and increased brightness, in both pulsed\cite{phillipsRaman} and continuous\cite{robinsRaman} configurations. Studies have shown that unpumped RF atom lasers can be coupled to waveguides \cite{guidedatomlaser}, and are first and second order
coherent, with Fourier limited linewidths \cite{esslingerlinewidth,g2atomlaser}.

The development of the cw optical laser was a significant advancement over the first pulsed ruby laser. The cw optical laser opened up many applications. The atom laser is a very promising source for both precision measurement and fundamental physics. Guided by the history of optical lasers, we describe here the first atom laser that is  simultaneously pumped and output-coupled to produce a free beam. These are important steps along the path to a continuous atom laser.
 The replenishment process can be divided into two critical components: a delivery system for filling an atomic reservoir with ultra-cold atoms, and a pumping mechanism for irreversibly and continuously transferring atoms from the reservoir to the laser mode.  The technical requirements on both parts of the replenishment system are stringent.  Nonetheless, recent experiments have demonstrated that a {\em delivery} system for atoms is feasible and possible.  
Chikkatur et al. showed that Bose condensed atoms could be periodically transported over large distances using a moving optical dipole trap \cite{ketterleexp}.  Further experiments with transport, based on interference of two counter-propagating lasers, have shown that dipole trapping techniques could be extended to provide continuous delivery of atoms \cite{Schmid}.  Magnetic guiding systems for ultra-cold atoms may also provide a path to future delivery systems \cite{david, greiner, greiner1, david1}.

The realisation of the pumping mechanism for a continuous atom laser has proved more
problematic. There are four critical requirements that are difficult to satisfy experimentally. The atoms should enter the laser mode {\em continuously} and {\em coherently}, that is, with the phase and
amplitude of the lasing condensate. Thus, atoms must make a transition that is Bose stimulated by the atomic lasing mode. The second requirement is that the pumping process is irreversible. It requires coupling to a reservoir. There are two reservoirs available, the empty modes of the electromagnetic field accessible via a transition from an excited atomic state, or the empty modes of the atomic field accessible via evaporation.  For a high coherence atom laser, the lasing mode must be a pure condensate making evaporation an unlikely mechanism for a highly coherent {\em continuous} atom laser. The third requirement is that the pumping system must be compatible with a continuous replenishment mechanism. This suggests strongly that there be a {\em physical} separation between the source and the lasing condensates. A physical separation with a stimulated transition between the source and the lasing mode isolates the lasing mode from phase kicks and heating that would result either as necessary consequence of the replenishment system ( for example in the replenishment system demonstrated by Chikatur et al. where condensates are merged \cite{ketterleexp}) or as a consequence of an imperfect delivery system. Finally, the fourth condition on a pumping system is that it should be possible to continuously output-couple atoms from the laser mode into a beam, while the pumping mechanism is operating. 

A number of previous experiments observing the process of superradiant Rayleigh scattering appear to offer a physical mechanism for providing pumping through  matter wave amplification \cite{kozuma,inouye}.  Superradiant Rayleigh scattering occurs when a far off resonant laser illuminates an elongated BEC.  A matter wave grating forms along the long axis of the BEC and atoms are preferentially scattered into non-stationary momentum states.  By providing a moving 'seed' in the $|2\hbar k\rangle$ momentum state researchers were able to demonstrate pulsed coherent amplification via the Rayleigh superradiance mechanism.
However, this type of matter-wave applification is a transient phenomena, observed over timescales ranging from tens \cite{inouye} to hundreds of microseconds \cite{kozuma}. On longer timescales, scattering into successively higher momentum modes appears unavoidable \cite{zobay,dan_comment}, resulting in the well known 'fan' shaped scattering pattern \cite{inouye2}.

Two promising mechanisms for providing a pumping mechanism consistent with a continuous atom laser have been recently demonstrated.  The first is {\em far detuned} stimulated
Raman scattering \cite{Schneble,Yoshikawa}, in which atoms in one internal atomic state are Bose-stimulated to make transitions into an alternate atomic state. The second, reported by Ginsberg et al. \cite{Hau}, is a {\em resonant} coupling driven by electromagnetically induced transparency (EIT), demonstrated as stimulated decay of atom pulses into a condensate in a freely falling frame in the context of quantum information processing. 
In both cases the coupling from the source mode is irreversible,
and the laser mode is dark to the photons produced by the stimulated transitions.

In this Letter, we demonstrate the realization of a pumped atom laser that fulfills the four
criteria listed above.  We demonstrate the pumping through measurements of the source and laser-mode atom number, making a rate equation study of the pumping process. Using the technique of sympathetic cooling pioneered by Myatt et al \cite{spinor}, we produce two independent, spatially separated clouds of ultra-cold atoms in different internal atomic states, thus creating the source and laser modes.  The source is coupled to the laser mode via a three-step process: radio-frequency coupling to an appropriate magnetic state; excitation by an optical field that is resonant only with the source atoms; and finally stimulated transitions from an excited atomic state into the laser mode \cite{joseph,santos}.   
 
 The linewidth of a pumped {\em optical} laser can be made dramatically lower than the linewidth of a bare cavity with a short lifetime \cite{wallsmilburn}.
 In addition, density fluctuations in an atom laser may be suppressed if the laser is operated at high density with significant three body recombination in analogy to intensity noise suppression in an optical laser \cite{simon}. Operating in this regime may come at the cost of heating and loss of coherence. There are many open questions about the operation of a continuous atom laser. Despite a large body of theoretical work \cite{craigreview}, it is as yet unclear what the classical and quantum properties of a pumped atom laser will be because the theories have yet to be applied to a specific system.

The pumped atom laser is an open quantum system. The appropriate 
description is through the master equation that governs the evolution 
of the reduced density matrix for the system in contact with the 
environment \cite{wallsmilburn}:
\begin{eqnarray}
\dot{\hat{\rho}}_{R} &=& \hat{\mathcal{L}}\hat{\rho}_{R} = -\frac
{\imath}{\hbar} \big[ \hat{H}, \hat{\rho}_{R} \big] + \hat{\mathcal
{K}} \hat{\rho}_{R},\\
\hat{H} &= &\hat{H}_\text{a} + \hat{H}_\text{lp} + \hat{H}_\text{a-lp}.\nonumber
\end{eqnarray}
The first term on the right hand side, the commutator of the system 
Hamiltonian with the reduced (system) density operator, governs the 
coherent dynamics of the system. For our system, $\hat{H}$ includes 
the Hamiltonian for the atoms $\hat{H}_\text{a}$,  the Hamiltonian for the local 
photon modes inside the lower condensate  $\hat{H}_\text{lp}$, and 
the Hamiltonian for the coupling between the atoms and the local photon modes $\hat{H}_\text{a-lp}$. 
The second term on the right hand side, $\hat{\mathcal{K}}$,  is the Liouvillian describing both spontaneous 
decay from an excited state and the decay of the local photon 
modes into the surrounding vacuum modes. In the case of evaporative 
pumping, the Liouvillian $\hat{\mathcal{K}}$ describes the 
collisional coupling of the atoms to the empty modes of the atomic 
field. In both cases, the irreversible step essential to pumping is 
provided by coupling to one of only two available reservoirs 
(depending on the design of the experiment) and is described by the 
Liouvillian on the right hand side of the master equation above.

In order to produce a pumped atom laser, we first load a three beam retro-reflected ultra-high vacuum magneto-optical trap (MOT), background pressure $10^{-11}$ Torr, with more than $10^{10}$ atoms of $^{87}$Rb with a cold atomic beam derived from a laser cooled atomic funnel. 
After loading, we compress the MOT and subsequently switch off all confining magnetic fields.
Through polarization gradient cooling we achieve temperatures around 40\,$\mu$K with little loss in number.
The sample is then subject to an intense, approximately 20\,$\mu$s, burst of $\sigma^-$ light tuned to the $|F=2\rangle \rightarrow |F'=2\rangle$ D2 transition which optically pumps the atoms predominantly into the magnetically trapped $|F=1,m_{\rm F}=-1\rangle$ ground state.  By precise control of this pulse time (on the order of 0.1\,$\mu$s) we also create a small sample of up to $5\cdot 10^6$ magnetically trapped $|2,2\rangle$ atoms. 
Subsequently, a magnetic quadrupole field is switched on to 200\,G/cm in less than 10\,$\mu$s confining both the $|1,-1\rangle$ and $|2,2\rangle$ atoms.  The atoms are transported over 20\,cm from the MOT chamber by moving the trapping coils with a mechanical translation stage, after which they are transferred into a harmonic magnetic trap.
The temperature of the atoms is further reduced using forced radio frequency evaporation.
Due to their larger magnetic moment, the $|2,2\rangle$ atoms are more tightly confined in the magnetic field than the $|1,-1\rangle$ atoms, and hence the evaporation does not directly cool them.
They are, however, sympathetically cooled via elastic collisions with the $|1,-1\rangle$ atoms \cite{spinor}. 
We arrange the population ratio of the $|1,-1\rangle$ and $|2,2\rangle$ atoms such that they  undergo the phase transition to BEC at approximately the same temperature. Via this method, we produce Bose condensed samples with up to $10^6$ atoms in each state \cite{comment2}.
The magnetic trapping parameters are $\omega_r=2\pi\cdot130$\,Hz and  $\omega_z=2\pi\cdot13$\,Hz for the $|1,-1\rangle$ state at $B_0=2$\,G, leading to a Thomas-Fermi radius of each cloud of approximately 5\,$\mu$m. The different magnetic moments of the two clouds lead to a gravitational sag between their centers of 8\,$\mu$m.    Hence, the two clouds of atoms overlap only slightly, with the $|2,2\rangle$ source condensate located above the $|1,-1\rangle$ laser mode condensate (Fig. 1a).  

In order to measure the effect of pumping, it is essential that the number of atoms in each state is stable from one experimental run to the next one. For this purpose, we have refined many details of the apparatus, including very efficient baffling against stray resonant light, very low uncertainty and drift in laser frequency, intensity  and polarization, and good vibrational and thermal stability of the trap. We measure the statistical error in the atom number by taking a statistical ensemble and calculating the standard deviation. There is no slow, systematic drift in the measurements, and consecutive measurements are independent of one another. This allows us to divide the measured standard deviation by the square root of the number of data points to get the error in the average number. For a data set comprising 20 measurements, the error in the mean number of atoms in each state is as low as 1\%.

Having prepared the atomic samples, we set up conditions for pumping as illustrated in Figure 1. Starting with an initial number of atoms in each state (Fig.1b), we apply a weak continuous radio frequency field to the upper source cloud which couples atoms from the $|2,2\rangle$ state, through $|2,1\rangle$ to the $|2,0\rangle$ state.  This coupling is highly spatially selective and does not effect the $|1,-1\rangle$ cloud.  These atoms begin to fall away from the $|2,2\rangle$ source cloud (Fig.1c).   Simultaneously we apply approximately 10 picowatts of upward propagating $\pi$ polarized light resonant with the $|F=2\rangle\rightarrow |F'=1\rangle$ transition.
Although this light is resonant in energy with the $|2,2\rangle$ source atoms, they are prevented from absorbing photons by atomic selection rules. Hence, the source cloud is unaffected by the pumping light.  For pumping the laser mode, we target the $|2,0\rangle$ atoms (Fig.1d).  
As these atoms fall they may make a transition into the excited $| F'=1,m_{\rm F}=0\rangle$ state from which they are stimulated to emit into the laser mode $|1,-1\rangle$ by the atoms already present in that mode. The $\sigma$ photon emitted in this process carries the phase difference between the pump atoms and the condensate.  
 Finally, the $|1,-1\rangle$ laser mode atoms are output-coupled to produce the atom laser beam in the $|1,0\rangle$ state (Fig.1f). An absorption image of the ultra-cold atoms used to build the pumped atom laser system is shown in Figure 1g.  A comparison of our experiment with previous work on Raman superradient scattering and EIT is given in the additional material.

We need to ensure that the momentum of the pump-atoms can be matched to the BEC. This means that the atomic velocity after the emission of the photon has to lie within the velocity spread of the BEC.
We have arranged for the magnetic trapping frequencies not only to position the two clouds as close as possible to one another, without significant spatial overlap, but also such that the velocity acquired by a $|2,0\rangle$ atom in falling from the center of the $|2,2\rangle$ cloud to the center of the $|1,-1\rangle$ laser mode (12\,mm/s) can be canceled by the absorption of an upward propagating $\pi$ photon and the emission of an appropriately directed and phased $\sigma$ photon; a single photon recoil corresponds to 6\,mm/s.  
The velocity at the laser mode center can be tuned by around $\pm 2$\,mm/s by moving the coupling surface within the source cloud up or down \cite{comment7}.
While the pump atoms are falling through the $|1,-1\rangle$ laser mode, the velocity varies by $\pm 3$\,mm/s due to gravity and the time for which the pumping atoms satisfy momentum resonance with the laser mode is much shorter ($\sim$ 100\,$\mu$s) than the traversal time across the laser mode ($\sim$ 1\,ms). 
The velocity spread of the laser mode is on the order of 0.3\,mm/s, 
thus, canceling the atomic momentum of the $|2,0\rangle$ state requires an extreme level of control over pumping parameters.

We take blue detuned absorption images of the expanded source and laser mode clouds using the Stern-Gerlach effect to separate the two internal states;
applying a magnetic field gradient during the trap switch off, the atoms experience a magnetic moment dependent force.
The resulting difference in velocity of the $|1,-1\rangle$ and $|2,2\rangle$ atoms causes the two clouds to move apart.
After 17\,ms of free fall, they are well separated, and we take an absorption image with a 100\,$\mu$s light pulse resonant with the $|2\rangle \rightarrow |3\rangle$ transition. 
Just prior to the imaging pulse, we pump all of the F=1 atoms into the F=2 state using a 200\,$\mu$s pulse of $|F=1\rangle \rightarrow |F'=2\rangle$ light. The short time scale ensures that the re-pumping process does not distort the image due to momentum diffusion \cite{comment3}.  Our images of pure condensates do not have the standard Thomas-Fermi form due to a combination of lensing effects due to blue detuning and aberrations of our imaging system.  We have verified that the experiments described in this paper are performed with near pure $|F=1,m_F=-1\rangle$ condensates.  

In order to isolate and study the pumping mechanism we run the atom laser system without producing a $|1,0\rangle$ output beam, in effect studying a continuously pumped BEC. Figure 2 demonstrates the effect 200\,ms of pumping has on the $|1,-1\rangle$ condensate; the left hand image is taken 200\,ms after production of the mixed sample without the $|2\rangle\rightarrow |1\rangle$ pumping light and without the RF coupling from the $|2,2\rangle$ source. The absorption image shows the unpumped F=1 laser mode BEC (the lower cloud) with the F=2 source atoms above. The curves below are horizontal cross-sections through the absorption images showing the optical depth of each atom cloud. There are $(6.7\pm0.5)\cdot10^5$ atoms in the source and $(5\pm0.4)\cdot10^5$ in the laser mode. 
The central image shows the effect of 200\,ms of pumping on the laser mode BEC.  The source is almost completely depleted and the laser mode atom number has increased to $(7.2\pm0.4)\cdot10^5$.  The third column shows the difference between the pumped and unpumped images.  The cross-section through the spatial distribution of the F=1 atoms displays a flat topped profile, proving that the increase in atom number is due predominantly to growth in the (parabolically distributed) condensate fraction.  The wings on the edge of the difference image stem from a small increase in the Gaussian thermal component.
The pumping efficiency, which we define to be the growth of the laser mode compared to the loss from the source, is typically 15-20\% and depends critically on the pump light intensity, the F=2 output-coupling strength and position, and the inital number of atoms in the F=1 BEC and F=2 source.  The quantitative dependence of the pumping mechanism on these parameters is an important area of research, both theoretically and experimentally, but is beyond the scope of this work. For the data presented in Fig.2 we measure a pumping efficiency of $(35 \pm 10)\%$.  

In order to create a pumped atom laser, we simultaneously run the pumping mechanism while output-coupling from the F=1 BEC.   This allows the atom laser mode to be simultaneously replenished and depleted. 
In our experiment, the condensate and source atom numbers and the duration of the pumping dictates that we produce a low output flux in the atom laser.  This prevents a direct measurement of the number of atoms in the output beam through absorption imaging.
Instead, we infer an increase in flux by measuring the atoms remaining in the laser mode BEC after simultaneously pumping and output-coupling. 
We take four consecutive sets of data for each 200\,ms run of the atom laser; the initial atomic samples (Fig.3a), the pumping only (Fig.3b), the laser mode output coupling only (Fig.3d), and the full pumped/output-coupled system (Fig.3c).  While pumping or output-coupling independently the laser mode increases or decreases, respectively. Running the full system with appropriately adjusted experimental conditions leaves the population of the laser mode essentially unchanged.  We can be confident that the laser mode {\em is} being replenished because the output-coupling mechanism is independent of the pumping.  The RF transitions it induces cannot be turned off by the pumping, and hence for the laser mode to remain constant new atoms must have been added.   

We use a simplified rate equation model to verify that output-coupling and pumping are independent from one another.
     $\dot{N}_2 = - \gamma_{p}  N_2$ describes the change in the population of the source BEC.  
     It is independent of the population of the laser mode BEC and comprises all output-coupling and loss mechanisms. From the depletion of the source (not shown in Fig.~3), we measure $\gamma_{p} = 9$s$^{-1}$.  
We omit the population of the intermediate pumping states.
$\dot{N}_1 = \gamma_{p}  N_2 \alpha  N_1    - \gamma_{o}  N_1$ describes the 
population of the laser mode BEC. 
The first term is the pumping rate, where $\alpha N_1<1$ is the pumping efficiency.
Finally, $\gamma_{o}N_1$ is the output-coupling rate into the atom laser beam. 
Using $\gamma_{o} = 1.4$s$^{-1}$ and $\alpha= 6.5 \cdot 10^{-7}$ 
we calculate the curves in Fig.3.
The fact that the three experimental data points can be reproduced by assuming two independent rates shows that the output-coupling indeed has no significantly adverse effect on the pumping mechanism. 
We calculate that we output-couple $8.3\cdot 10^4$ atoms into the atom laser beam in  the fully pumped system and $7.0\cdot 10^4$ atoms without pumping, thus increasing the laser beam population by 20\%.  

In our current apparatus, we cannot directly measure that the coherence of the pumped condensate is preserved.  The flux of stimulated photons from the pumping process is extremely low, around $10^5$ photons over 200ms, and we have not yet established a method to separate the incoming pump photons from the outgoing stimulated photons. It is also not currently possible for us to measure the output flux of the continuous atom laser directly due to the low density of the output coupled atoms and the field of view of the imaging system. If one could utilize the recently demonstrated technique of atom counting using ultra-high finesse optical cavities \cite{esslinger_revsci} it would be possible to measure not only the flux, but also the effect of pumping on the quantum statistics of the output beam, thus verifying the phase coherence of the pumping process.  Finally we note here that, in further experiments, we have been able to use a {\em thermal} cloud as the F=2 source of atoms for pumping.

In summary, apart from satisfying the four criteria for pumping (compatibility with replenishment, coherent, continuous, and irreversible), there are two critical features of our system for pumping the atom laser. The first is that, to a very good approximation, neither the source cloud or the laser mode can absorb the pumping light. The second is that the pumping process drives atoms irreversibly into the laser mode, and that mode is dark to the optical transition that populates it. 
The system that we have presented here represents an important step towards a continuous, pumped atom laser. We have produced a pumped atom laser, derived from the interaction between a source, a trapped BEC and a freely propagating beam.  

\section{Additional supporting material}
\subsection{Raman superradiant scattering and EIT}
The mechanism we have described for pumping is related to pulsed Raman 
supererradiance that has been studied previously \cite{Schneble,Yoshikawa}. In our work, in the lab frame, atoms are driven from the 
$|2,0\rangle$ untrapped state into the $|1, -1\rangle$ trapped state allowing us to 
continue pumping indefinitely.  The condensate that we pump is 
stationary in the lab frame.  Furthermore, our experiment is 
performed in a double condensate geometry with internal states that 
allow simultaneous pumping and outcoupling to produce a freely 
propagating atom laser beam from a simultaneously pumped condensate.

There is a second mechanism that we have considered that is related 
to the EIT mechanism described in the work by Ginsberg et al 
\cite{Hau}. Similar to the exisiting Raman superradiance 
experiments, their work was necessarily performed in a pulsed regime. 
In our experiment, again, the geometry and choice of internal states 
allow the possibility of performing a related experiment 
continuously. The condensates in such experiment must have some 
spatial overlap. The $|2, 0\rangle$ atoms are produced in the overlap region 
where the $\pi$ polarised pump beam is absorbed and the $\sigma$ beam 
is produced. Both beams travel in the same direction. The amplitude 
of the $\pi$ polarised beam decays in the overlap region. The amplitude 
of the $\sigma$ beam grows. The atoms follow the dark state and are 
pumped continuously to the lasing mode. 

\subsection{Pumping via evaporation}
We have also investigated a pumping mechanism based on the evaporation of atoms \cite{holland1,holland2}.  In this system, we produced a partially condensed cloud of rubidium atoms in the $|1,-1\rangle$ state.
Applying a monochromatic RF field resonant near the edge of the trapped cloud causes transitions from trapped to untrapped states, selectively removing the higher energy atoms from the thermal distribution.
The remaining atoms rethermalize at a lower temperature, causing the condensate fraction to grow in number \cite{ritter}.
Due to the nature of magnetic traps, the pumping surface has to lie below the BEC to achieve efficient cooling.
Simultaneously with the evaporation, we produced a beam from the trapped cloud using a second radio frequency field resonant at the center of the condensate, as shown in Fig.4a. Unfortunately, this field preferentially removes low energy atoms from the thermal cloud (``anti-evaporation''), thus heating the system and {\em decreasing} the number of atoms available to be pumped into the condensate.  Additionally, the output-coupling process does not  distinguish between atoms in the lasing mode (the BEC fraction) and atoms in the source (the thermal cloud). Thus, by producing an output beam, we also unavoidably deplete the source.  Clearly in this system the beam output coupling has a detrimental effect on the pumping.
Two additional points are worth mentioning about the output beam from such a system. Because the cloud of atoms is only partially condensed the coherence properties of the output will likely be well above $g_2 = 1$ \cite{g2atomlaser}.  The strong thermal component of the atom laser beam, similar to amplified spontaneous emission in an optical laser, would have to be filtered and rejected. Finally, the output beam must pass through the evaporation surface used to cool the sample and we have observed that a large fraction of the $|1,0\rangle$ beam is spin flipped back into the trapped $|1,-1\rangle$ state as well as into the anti-trapped $|1,1\rangle$ state (Fig.4b), again reducing the usefulness of this system in future experiments and making measurements of population changes in the trapped states unreliable.
We therefore conclude from our measurements that evaporation is not a suitable method for pumping {\em the continuous atom laser} in a trapped atomic cloud, and that the source of atoms must be in a different internal state to provide for pumping through stimulated emission. 
Finally, having the source and the BEC in thermal contact adds a significant challenge to the problem of replenishment of the source since heating in either system will remove atoms from the BEC. A physical separation, however, is only possible in a pumping scheme that does not rely on re-thermalization through collisions.

\clearpage
\begin{figure}[ht]
\centerline{\scalebox{.6}{\includegraphics{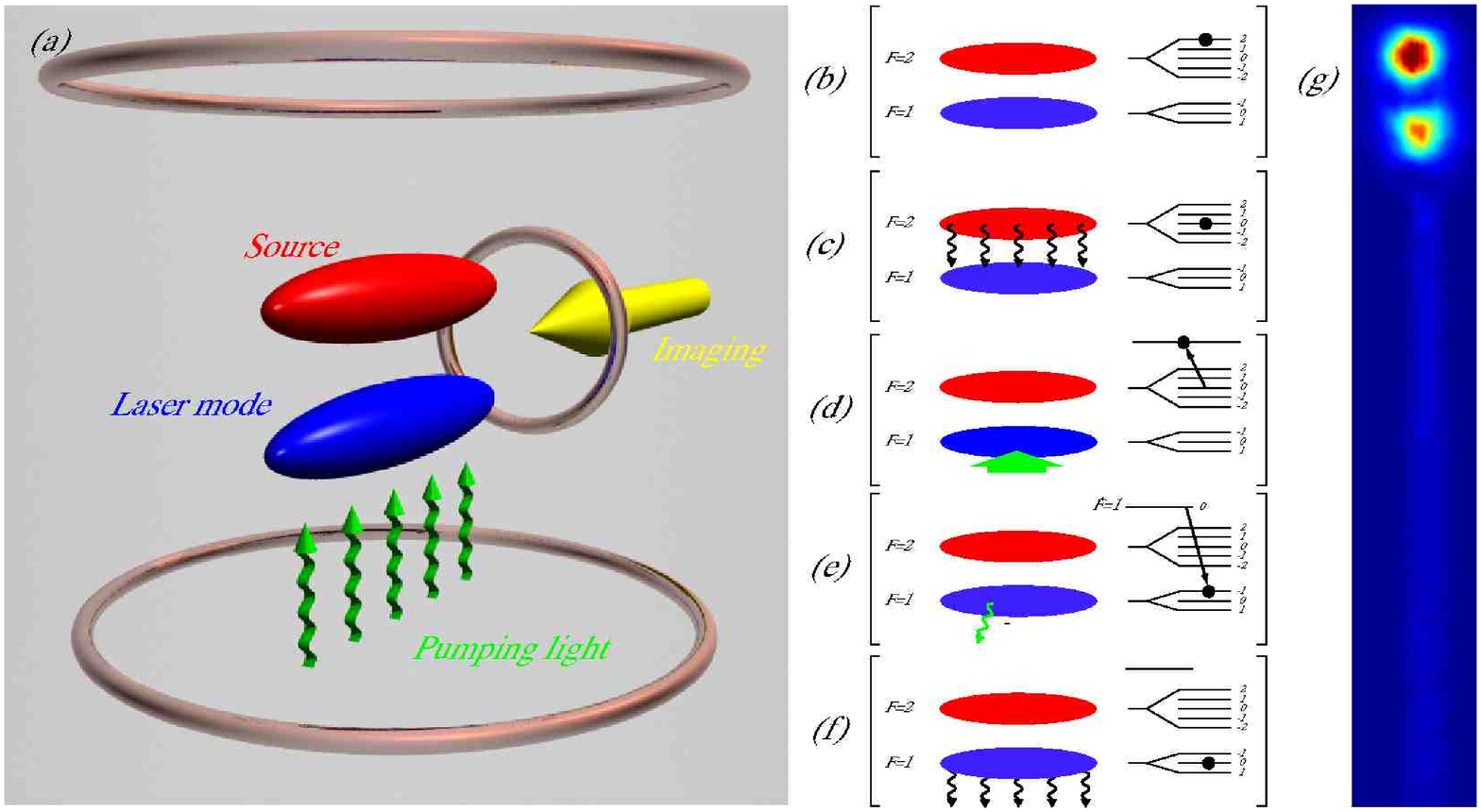}}}
	\label{fig.schematic}
	\caption{Schematic of the experiment (a) and pumping steps (b-f).  A radio frequency field spin-flips the atoms to the $|2,0\rangle$ state (b), and they fall under gravity (c). The light field couples the atoms to the F$'$=1 excited state from which they are stimulated to emit into the $|1,-1\rangle$ BEC. The atomic momentum is canceled from the absorption and emission of the photons (d) and (e).  A second radio frequency field finally output-couples the atoms into the $|1,0\rangle$ atom laser (f). (g) Absorption image of the experimental system, showing source, laser mode and output beam.}
\end{figure}

\clearpage

\begin{figure}[ht]
\centerline{\scalebox{.7}{\includegraphics{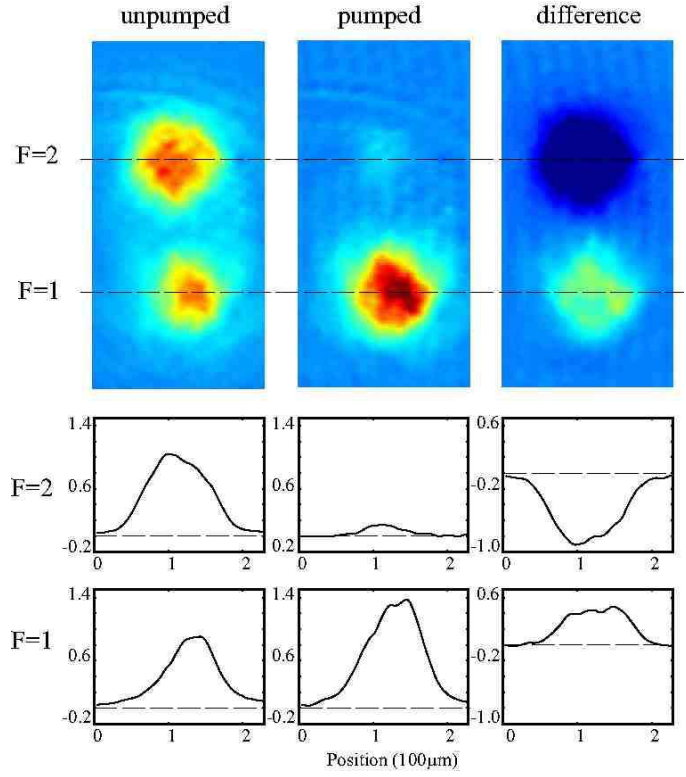}}}
	\label{fig.pumped_BEC}
	\caption{Pumping of a BEC. Blue detuned absorption images averaged over three identical runs of the experiment (top row), detuning of the probe laser from resonance is 7\,MHz.  The graphs below are horizontal cross sections through the absorption images, showing optical depth, averaged over 50\,$\mu$m in the vertical direction. The three columns correspond to: pumping off (left); pumping on (center); difference between pumped and unpumped (right).}
\end{figure}

\clearpage




\begin{figure}[ht]
\centerline{\scalebox{.7}{\includegraphics{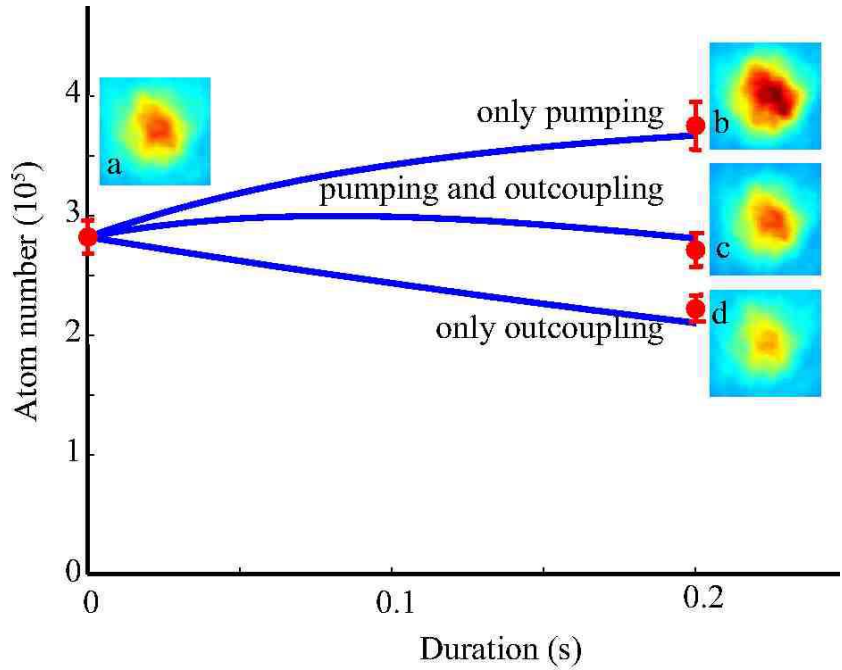}}}
	\label{fig.pumped_AL}
	\caption{A pumped atom laser. Atoms remaining in the laser mode after running the output coupler and/or the pumping. (a) No pumping or output-coupling (unpumped BEC), (b) only pumping (pumped BEC), (c) pumping and output-coupling (pumped atom laser), (d) only output-coupling (unpumped atom laser). The curves are calculated from the rate equation model discussed in the text.}
	\end{figure}
	
	\begin{figure}[ht]
\centerline{\scalebox{.7}{\includegraphics{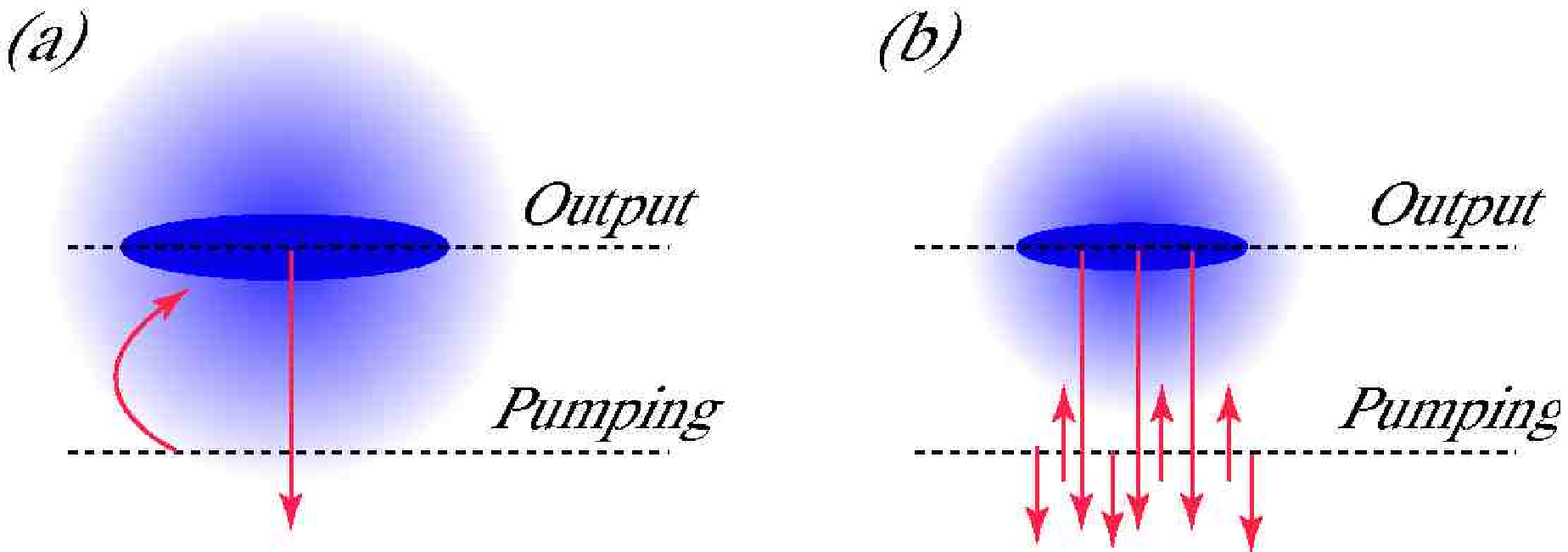}}}
	\label{fig.evap_pumping}
	\caption{Illustration of pumping through evaporation. This method breaks down because the output-coupling removes pump atoms, heats the system and the pump RF surface spin-flips atoms back into trapped states. }
\end{figure}


\begin{thebibliography}{99}
\bibitem{david1} Indeed, it may be possible to Bose-condense the transverse degrees of freedom of atoms continuously in a wave-guide, providing an alternate path to the continuous atom laser.  See for example the ongoing experiments in the group of D. Guéry-Odelin in Paris.
\bibitem{dan_comment}It may be possible to suppress this coupling to higher momentum modes, but further experimental work is required in order to confirm this.
\bibitem{comment2}The total number of atoms in each state can be adjusted independently (within limits) by changing the parameters of the optical pumping and the evaporative cooling.
\bibitem{comment7}The tuning range is limited by the fact that only when coupling close to the centre can the BEC can be effectively depleted.
\bibitem{comment3}The $|2,1\rangle$ atoms have the same magnetic moment as the $|1,-1\rangle$ atoms and are thus indistinguishable in our images. We omit the re-pump pulse to image only atoms in the F=2 state and correct the measured population of the $|2,1\rangle$ state for this background.
\end{thebibliography}
\end{document}